# Competency of the Developmental Layer Alters Evolutionary Dynamics in an Artificial Embryogeny Model of Morphogenesis

Lakshwin Shreesha[1] and Michael Levin[2,*]

1   Université Paris Cité
    UFR Fundamental and Biomedical Sciences
    Paris, France

2   Allen Discovery Center
    Tufts University
    Medford, MA

* Author for correspondence:
        Email: michael.levin@tufts.edu
        Tel.: (617) 627-6161

**Keywords:**      artificial life, in silico, artificial embryogeny, evolutionary computation, development, morphogenesis, basal cognition

# Abstract

Biological genotypes do not code directly for phenotypes; developmental physiology is the control layer that separates genomes from capacities ascertained by selection. A key aspect is competency, as cells are not a passive material but descendants of unicellular organisms with complex context-sensitive capabilities. We used an evolutionary simulation in the context of minimal artificial embryogeny to probe the effects of different degrees of cellular competency on evolutionary dynamics. Virtual embryos consisted of a single axis of positional information values provided by cells' genomes, operated upon by an evolutionary cycle in which embryos' fitness was proportional to monotonicity of the axial gradient. Evolutionary dynamics were evaluated in two modes: hardwired "mosaic" development (genotype directly encodes phenotype), and a more realistic mode in which cells interact prior to evaluation by the fitness function ("regulative" development). Even minimal competency with respect to improving their position in the embryo results in better performance of the evolutionary search. Crucially, we observed that as competency of cells masks the raw fitness of the genomes, the phenotypic fitness gains are then mostly due to improvements of cells' developmental problem-solving capacities, not the structural genome. This suggests the existence of a powerful ratchet mechanism: evolution progressively becomes locked in to improvements in the intelligence of its agential substrate, with reduced pressure on the structural genome. A feedback loop in which evolution increasingly puts more effort into the developmental software than perfecting the hardware explains the very puzzling divergence of genome from anatomy in species like planaria, identifies a possible drive for scaling intelligence over time, and suggests strategies for engineering novel systems in silico and in bioengineering.

# Introduction

The canonical picture of evolution is that of a process in which genomes change over time due to selection of organisms' functionality in some environmental niche in which they compete. One of the critical aspects of real biology however, which is only sometimes taken into account in evolutionary computation and theoretical biology efforts, is that the mapping between genotype and phenotype is not direct [1-13]. The genome is not a direct blueprint of the anatomy, and genes generally do not directly encode for structure and function of the organism. The layer of control that sits between genomes (on which mutation operates) and anatomy (the phenotype which is the subject of selection) is developmental physiology. Real organisms emerge as the results of a complex set of interactions among cells, with anatomical order and functionality being the result of cellular activities. While genomes specify the cellular hardware (proteins), it is the software studied by developmental biologists that is ultimately responsible for the organism's overall structure and behavior.

The indirect relationship between genotype and phenotype has a number of important implications. For example, it is currently impossible to guess the anatomy of an organism by examining its genome – overall symmetry type, number and kinds of organs, size, regenerative capacity, etc. can only be estimated if one compares a genome to that of another organism for which all of these things are already known. Likewise, if one has access to complete genomes, for example of the frog and axolotl, one cannot guess the shape of a chimeric embryo: will a "frogolotl", consisting of 50% of each kind of cells, make legs (like an axolotl larva) or not (like a tadpole)? This is because while much research has shed light on molecular mechanisms necessary for morphogenesis, the field still largely lacks an understanding of the key dynamics that determine form and function: large-scale anatomical decision-making by cellular collectives [14-16].

Moreover, the simple story of genomes determining anatomy is shown to be incomplete by examples such as the highly regenerative planarian [14]: due to some flatworm species' reproduction by fissioning and regeneration, they retain mutations made in the parents' body and pass them to the offspring (somatic inheritance). As a result, planaria have an incredibly messy genome (indeed, worms are mixoploid – different numbers of chromosomes in each cell) but have the most reliable anatomy: every fragment of a planarian regenerates a perfect worm each time. They are essentially immortal, and highly resistant to cancer [17]. How can the animal with the most chaotic genome have the most reliable, robust anatomy? Fundamental knowledge gaps in this area concern not only basic evolutionary developmental biology but also have practical implications in that they limit our ability to make desired system-level changes to complex anatomy in the context of regenerative medicine [18-20].

Developmental outcomes are not micromanaged by the genome; instead, they result from the interactions of subunits (cells) that were once independent organisms – they are an agential [21], not a passive, material. One of the most remarkable qualities of this process is its competency in reaching an adaptive outcome despite novel starting states and perturbations [15,22,23]. For example, mammalian embryos can be split into pieces, and each piece gives rise to a complete organism (monozygotic twinning). Some animals keep these regenerative capacities into adulthood – salamanders whose limbs (or eyes, jaws, tails, etc.) are amputated will re-grow exactly the missing portion and then stop when the correct structure is complete [24]. Tadpoles with scrambled faces become largely normal frogs, as the craniofacial organs move in novel paths until the correct configuration is achieved [25-27]. This means that mutations resulting in noise or changes in initial positions of the organs, which would have been disastrous for a hardwired architecture, will not have a strong effect on survival because the competency of the tissues will make needed reconfigurations despite the errors in initial state. And finally, skin cells isolated from frog embryos self-assemble into a novel, coherent, motile proto-organism that has properties like kinematic self-replication [28-30].

The competency of the developmental layer has been proposed to result from the navigation policies of a collective intelligence of cells in anatomical morphospace, as an evolutionary precursor to the intelligence of neural cells which are well known to navigate 3-dimensional and other problem spaces [15,22]. What is very poorly understood are the implications of these dynamics for the evolutionary process. How do diverse levels of competency in the cellular collective impact the rate and course of the evolutionary process?

We undertook a quantitative investigation of these questions using a minimal model of artificial embryogeny. In our system, virtual embryos consist of a 1-dimensional array of integers, with fitness being proportional to the degree of monotonicity of those values. This simulates an animal with a single axis of positional information values (such as the anterior-posterior axis) [31,32]. The values of an animal's axis (phenotype) are determined by its genome, which sets the order. In the baseline (control) case, we use a direct encoding where the phenotype is a direct consequence of the genotype – the genes directly specify the order of values in each animal. Under these conditions, a genetic algorithm eventually produces genomes in which all the values are in the correct (monotonic) order. We compare these outcomes to a more realistic case, in which the mapping is not direct: during the developmental process, the individual cells have some degree of competency to rearrange themselves based on their local environment. Cells can move to numerically more-advantageous positions, before fitness evaluation (corresponding to embryogenesis in which cells act before the animal's adult fitness is ascertained in the environment). This in effect performs a limited "bubble sort" [33], highlighting the conceptual similarity between sorting algorithms and navigation in a geometric problem space. Importantly, our system does not include Lamarckian inheritance (it features a strong barrier between soma and germline): the rearrangements are present for each individual but the only thing that gets passed on to their offspring is their un-altered (pre-swap) genome. In other words, the results of enacted competency are not passed on transgenerationally [34,35].

We varied the degree of competency (how many swaps are cells allowed to perform before fitness evaluation), and tracked the dynamics of the resulting evolution, both in terms of raw genotypic fitness (how fit would the individual be if its genome were allowed to determine the phenotype "as-is" – with no cellular movement to improve ordering), and phenotypic fitness (how good is the axis order after the cells have finished their movements).

We observed a number of interesting outcomes. First, including a developmental layer improves evolutionary efficiency, proportion to the degree of the cells' competency. Second, in mixed populations, competent individuals tend to eventually dominate the population. Third, when the degree of competency is itself allowed to evolve, populations settle on a specific (sub-maximal) level of competency. Finally, and most critically, we observed that because competency hides genetic deficiencies from selection, pressure to improve the structural genome is released, while pressure to improve the competency of cells is increased. These dynamics establish a positive feedback loop in which populations advance by progressively improving cellular capacities, not just starting configurations. This provides an explanation for the otherwise mysterious disconnect between planarian genomes and their amazing anatomical robustness, and suggests the existence of an evolutionary ratchet working to optimize intelligence in even very basal forms (i.e., the collective intelligence of cell groups) [36-41].

# Methods

General computational details
- All code was written in python 3.9 and is available on Github: [https://github.com/Niwhskal/CellularCompetency]
- The numpy python package was used for creating and manipulating arrays.
- Seaborn, and matplotlib python plotting libraries were used for the plots.
- Experiments were run on an ARM based M1 13-inch 8 core macbook pro (2020), equipped with 8GB of RAM.
- Experiments were run as separate processes using the multiprocessing python package. At any given time, eight different processes were initialized to run eight different experiments. Each of these were assigned to one of eight CPU cores. A single process was initialized to run a single thread.
- Version control was ensured through the Git version control system.

Creating Populations for Evolution

A population consists of a number of individuals. Each individual is considered to be a one-dimensional array of fixed size. Each byte of this array is considered analogous to the cell of an individual. Each cell is initialized with a single integer from the interval [1, array_size] (two or more cells are allowed to carry the same integer value). The integer-value that a cell carries is considered to be a gene, and its magnitude as a specification of how much it is expressed. We call the set of all genes of a population as the individuals' "Structural genes". In all our experiments an array_size of 50 is chosen. Unless otherwise mentioned, each population is initialized with 100 individuals.

Fitness of an Individual

The fitness function specifies the objective to our genetic algorithm. We define the fitness of an individual as the degree of order amongst its genomes, i.e., the degree of order among the bytes of an array. If an individual has its genes in ascending order by value, then we attribute it to have a fitness of 1.0 (maximum), if an individual's genes are randomly ordered, we attribute it to have a fitness of 0.0 (minimum). We calculate the fitness (the degree of order) of an array by counting the number of non-inversions prevalent in it. This count is normalized to span the range [0, 1]. Formally,

Let an array, A be initialized with a set of n values. Let A(0), A(1), A(2),... A(n) be its elements. Now, if i < j and A(i) > A(j), then the pair of indices (i,j) is called an inversion of A. To get all non-inversions of the array, we execute the following:

$$nonInv(A) = \#\{(A(i), A(j)) | i < j \text{ and } A(i) < A(j)\}$$ where $i \neq j$ and $i = 0 \ldots n-1, j = 1 \ldots n$

Where '#' = number of elements.
This count is normalized as follows:

$$c = \frac{nonInv(A)}{nC2}$$

finally the fitness is reported on an exponential scale:
Fitness $= 9e^{-c}$
We choose an exponential scale in order to zoom-in to higher fitness values.

Hardwired and Competent population

In order to test the role of competency in evolution, we define two sets of populations to evolve. One of these is the "Hardwired population" and the other is the "Competent

Population". A key differentiating factor between the two is the ability of the competent individuals to rearrange their cells. Consider two individuals in a particular generation: C1 from the competent population, and H1 from the Hardwired population. C1's cells have the phenotypic-capability to rearrange themselves by assessing their position relative to their neighbors. In contrast, H1's cells do not have this trait (i.e., H1's genome is hardwired). The result is that C1 can move its cells around to increase its fitness over the course of a generation, whereas H1 will always have the same fitness from birth to death.

Competency

Competent individuals are made of cells which are able to move around, with each cell working to improve the monotonicity of its local microenvironment. To do so, each cell of a competent individual looks at its right neighbor cell's value to check if it is out-of-order. If so, it swaps its value. The swap is considered if and only if it increases fitness of the individual. In case the cell in question already finds itself in order with respect to its neighbors, then no swap occurs and we move onto the next cell. This process occurs sequentially from index 0 of the array to index N. Formally:

swap $array[i]$ with $array[i+1]$, if $array[i] > array[i+1]$

In order to prevent each cell from swapping without limit, we add constraints on how much it can swap. These constraints are elaborated in the Results Section.

Pseudocode for this process is as described below:

Require: Competent_population (int), sc: user defined limit on swaps (int), ff: function to calculate fitness, nif: function to calculate number of non-inversions, cfs: function to check if a swap results in a fitness increase.

For individual in competent_population:
    total_swaps_available = nif(individual)
    defecit_swaps = total_swaps_available - sc
    swapsToUse = choose total_swaps if deficit is -ve or SC if + ve
    counter = 0
    not_swapped = 0
    **while counter < swapsToUse and not_swapped ≤ size(individual) −2:**
        for cell_index in individual:
            if individual[cell_index] > individual[cell_index +1]:
                if cfs(individual, cell_index):
                      swap and replace individual
                      counter ++
                else:
                      not_swapped ++
            else:
                not_swaped ++
        if counter > swapsToUse or not_swapped >size(individual)-2:
            break
    return individuals

Genetic algorithm

In order to evolve populations, we iteratively pass them through four different stages (Fig. 1):

1. Fitness Calculation: Fitness calculation occurs in slightly different ways based on the type of population used. A hardwired population has only one fitness: the hardwired/genomic fitness. However, in a Competent Population, there are two types of fitnesses calculated: the hardwired/genomic fitness and the competent/apparent fitness. The competent fitness is calculated after the individual has phenotypically swapped its genes around to boost its fitness.
2. Selection: This stage involves selecting for the fittest 10% of individuals in a population. This is done by calculating the fitness of all individuals in a population and choosing the top 10%. For a hardwired population, selection is based on its hardwired/genomic fitness. Whereas, in a competent population selection is based on the competent fitness
3. Selection in the competent population involves an important aspect. The competent populations have a pre-swap stage during which they are analogous to have been "just-born". Post this, they are bestowed with the ability to swap their genes (phenotypically) and increase their fitness. Once they complete swapping they are considered to be at the peak of their lives during which we assume selection to occur. Therefore, even though selection is based on their phenotypic fitness, it's the genomes of these individuals which are being propagated into the next generation.
4. Cross-Over: Post selection, the fittest 10% of a population remain. In order to repopulate it back to its original strength we carry out a process of reproduction called cross-over. It occurs as follows: Two individuals are involved, each of these are split at a random location along their length. One half of Individual 1 is swapped with the same half of Individual 2 to give rise to two children. Fig.1 contains an illustration of this process.
5. Mutation: Post Cross-over, a population is subjected to random mutations. Only point mutations occur, i.e., only a single structural gene of an individual in a population is set to a random value between [1, array_size]. We set the probability of an individual receiving a point mutation to be 0.6
Fitnesses of the best individuals in their respective population are plot over time (i.e., generations).

# Results

A minimal system for investigation of effects of cellular competency on evolution

Intelligence (problem-solving and learning) takes place on multiple levels in biology, including at the level of cells [42] and at the level of populations [43-45]. We sought to construct and analyze a minimal model in which developmental competency and evolution could be studied together, so that the fundamental dynamics were revealed as clearly as possible without the complexity of real pathways and frozen accidents of biological evolution. Thus, we built a virtual embryogeny model in which fitness was defined by the degree of monotonicity of a 1D array of numbers, simulating a minimal metazoan bodyplan – a single axis of positional information (Figure 1). The initial sequence of numbers for each organism was given by their genome. In "hardwired" individuals (Figure 1A), that sequence was exactly the sequence seen by the fitness function – their genome directly encoded their phenotype. We then implemented different degrees of competency during a developmental period in which cells were allowed some degree of movement relative to their neighbors (Figure 1B). The quality of the resulting gradient within each individual was evaluated after this period, enabling phenotypic fitness to diverge from raw, genotypic fitness (the fitness the animal would have had based purely on their genome), depending on how much cell movement was permitted. This corresponds to different degrees of capacity for cells to optimize homeostatically-preferred local conditions with respect to informational signals such as positional cues and polling of neighboring cell states [46-54]. An evolutionary cycle was implemented around these developmental events (Figure 1C), where the top individuals in each generation were used as the founders of the next generation [55,56]. This enabled study of the dynamics over time as a function of different degrees of cellular competency. In all plots, the shading around each line reports the variability across repeat runs of each experiment.

Developmental competency accelerates evolutionary search

We first compared the time-course of evolutionary search towards a fully-ordered axis in hardwired individuals (controls) vs. those with varying degrees of competency to rearrange their cells before selection, over 1000 generations (Figure 2). We found that after 500 generations, the hardwired population (with no swapping capability) had the least fitness. Table.1 provides a summary of the generation number at which each population crossed different fitness thresholds. We concluded that higher number of allowed swaps (greater competency) led to the best performance, with 100 swaps being already maximal. It was also seen that the variance between repeat runs of each experiment (N=10) decreased with larger competency, suggesting that more competent architectures are also more consistent in their performance over time. It is to be noted that hardwired individuals gradually improved to reach peak fitness, taking well over 1000 generations to do so, whereas the most competent individuals did so in under 100 generations. Figure. 2 is an indication of the role competency plays in non-linearly improving the rate of fitness of a population.

One specific example (competency = 100 swaps) is shown in Figure 3. Genotypic and phenotypic fitness of this population were similar for the first ~35 generations. But after that, the competent population experienced no more improvements in their genome (genotypic fitness flattened out). By this time, the genome was already so good that 100 swaps could convert each embryo to a very good phenotypic fitness. Thus, competency enabled excellent performance but reduced the pressure on the genome, so that raw fitness was much lower than that of a hardwired agent (whose only chance for improvement was via genetic changes).

Competent individuals take over mixed populations

Given these tradeoffs, we next asked how mixed populations of a total of 200 competent and hardwired individuals would evolve (Figure 4). We performed experiments varying the degree of competence, and the proportion of each type of individual in the initial generation.

At a low competency (20 swaps allowed per individual's developmental phase), the competent populations could dominate only when their numbers were same or much higher than those of hardwired individuals. As a minority (10%), they persisted but did not out-compete the hardwired ones. This was also true for competencies of 25 and 30 swaps. But, at a level of 31 swaps or above, they were able to dominate the population (although not drive the hardwired ones to extinction) within just a few generations, even when starting as a small minority of the population. Finally, we checked the competency at which the hardwired phenotype would be driven to extinction when the competent are a minority (Figure 5). We found that a competency of 100 swaps or greater was required to drive the hardwired individuals to extinction.

A summary of the relationship between competency values, ratio in which individuals were mixed and generation at which they dominated are shown in Table 2. We concluded that competency was a dominant strategy and even at very low prevalence, the population was quickly taken over by such individuals.

Evolution results in a high, constant level of competency

The prior experiments used hardwired values for the number-of-swaps available (competency) to the individuals of a population. In order to check what values would emerge during evolution where competency was itself an evolvable trait, we next set each individual's competency level (# of swaps permitted) to be determined by a gene in that individual's genome. During initialization, functional genes of all individuals were set randomly to low values in the range [0, 15]. However, during evolution we allowed each functional-gene to be mutated, potentially taking values across the range of [0, 400].

To determine which competency levels were preferred by evolution, we tracked competency gene values of the best individual over all 1000 generations, binned them into groups of 10, and then plotted the average value of each bin (Figure 6A). The prevalence of the competency allele rapidly rose, meandering during the evolutionary process. However, while evolution clearly explored the entire range of values (light green shading in 6A), it did not simply drive to maximize (400) competency – instead the optimal range was ~[250-300]. We provide a possible explanation for this surprising outcome in the Discussions section.

The results of evolving a competent population with variable competency over 1000 generations are shown in Figure 6B: the populations quickly settled at consistent configurations in which phenotypic fitness and genotypic fitness diverged considerably. This is a fascinating outcome because it suggests that a certain degree of competency reduces the pressure for improvements in the structural genome: once selection can no longer distinguish the quality of agents (because their high fitness could have been due to a great genome or to high competency repairing a terrible genome), it can only improve the population by increased competency, not by selecting better genetics.

To quantify this effect, we plotted the degree of correlation between genotypic and phenotypic fitness in these populations (Figure 6C), to determine how well selection (which sees phenotypic fitness only) can really judge genotypes (necessary to let the best individuals sire the next generation in a survival of the fittest scheme). We observed that correlation very quickly dropped (within about 10 generations), meandering a bit but centered around 0. We concluded that the presence of evolving competency strongly broke the ability of selection to choose the best genomes, and that the improvement process rapidly became dominated by individuals who succeeded because of their developmental competencies, not because of their continuously-improving genomes.

# Discussion

Evolution is driven by feedback between mutation and selection. However, these operate on two distinct aspects of individuals (genotype and phenotype), which are not in 1:1 direct correspondence [57-62]. The genome and local environment can be seen as the input layer, while anatomical outcomes are the output layer, of a complex computational process traversed by individuals during their lifetime. A complex layer of interactions (morphogenesis) sits between the specification of the cellular hardware (the genome) and the resulting form and function that provides adaptive fitness. In effect, development is a set of "hidden layers" that may provide the same advantages that hidden layers provide in artificial neural network architectures over simple perceptrons.

This developmental layer has been discussed in an evolutionary context among biologists [63-68], and generative encodings have been used in the artificial life/evolutionary algorithm communities [3,4,6-10,69-76]. Here, we focused on a specific and novel question: the implications, for the rate and course of evolution, of a developmental process that exhibits competency at problem-solving [22,23]. This is an essential question because evolution builds with agential, not passive, materials: cells were once independent organisms themselves, and they did not lose their competencies when multicellularity arose [36,37,42]. Instead, their ability to optimize specific variables, communicate with others, pursue goals in diverse problem spaces, learning capacity, etc. were harnessed toward larger-scale outcomes [77,78]. How might this affect evolvability and other aspects of the evolutionary process?

We produced a minimal simulation (Figure 1) that abstracted away many biological details to focus on a simple architecture: a phenotype easily optimized by traditional genetic algorithms, and a new component: competency of the individual cells to move, based on interactions with neighboring cells (simulating a single body axis morphogenetic gradient of positional information) [79-82]. The evolutionary algorithm basically performed a bubble sort [33], but with a new component: the individual elements of the set are not passive objects but have their own perspective and local goal (a specific relationship with neighbors based on their local properties, as cells are known to do [83-85]), and a limited competency to achieve that goal during their lifetime journey from genotype to judgement by fitness in their environment. Crucially, this was not a Lamarckian mechanism: each competent rearrangement was only in place "somatically" – for the lifetime of the organism, and did not pass on to its offspring, who inherited their parent's genome and would have to apply whatever competency it could anew.

<u>Limitations of the study</u>

Our framework was more complete than many evolutionary simulations because it included an explicit developmental layer between the genotype and phenotype. It was multiscale in the sense that there were actions taken by an evolutionary scale across individuals, but also by *components* of those individuals – the cells, which had their own perspective and local goals. However, our system clearly omitted a huge amount of biological detail with respect to cellular mechanisms of sensing, competition, cooperation, etc. Future work will add additional realism by including physiological layers, diverse cell types, and a multi-dimensional target morphology (e.g., 2D or 3D pattern instead of just one primary axis). There is also much that can be improved with respect to the specific mechanisms that cells use to implement their competency: a rich set of diverse genes will be added in the future to enable evolution to manipulate the different types of local goals and competencies. Moreover, recent discoveries in transgenerational inheritance [86-90] suggest that barrier between the genome and the phenotype is at least somewhat porous, and the effects of propagating sort order to offspring should be investigated. Fundamentally we explored a toy model virtual world in which the individual roles of selection and competency could be quantitatively dissected in the absence of confounding complexity – we sought generic laws [91-94], not a simulation of any one biological species.

The role of cellular competency in evolution

We found that providing cells with a minimal homeostatic competency with respect to improving their position in the virtual embryo results in better performance of the evolutionary search. Populations reach better fitness values, faster, when their cells are able to make up for deficiencies in their genomes (Figure 2). Indeed, in mixed populations, competent individuals tend to dominate and rapidly take over (Figure 4), as long as they have a minimal level of competency (Figure 5). The simulation highlighted the distinction between two properties of each individual that are often conflated or obscured in simulations that don't include an explicit competency step: genotypic vs. phenotypic fitness. Phenotypic fitness is the actual environmental performance that selection act upon. Genotypic fitness is the raw fitness the individual would have, if it had no competence; selection does not see genotypic fitness directly. Indeed, biology has many examples of evolution's attempts to gauge genomes that it cannot see directly, for example by fluctuating asymmetry [95,96] and the near universal standards of sexual selection for left-right symmetrical features (which in turn is correlated with lack of genetic damage) [97-99].

In this system, we see that competency results in good phenotypic fitness but takes pressure off of genotypic fitness, which settles at a sub-optimal level (Figure 3).

Perhaps the most interesting aspect was the role that competency plays in exacerbating the inability of selection to evaluate the genetic material that gets passed on to subsequent generations. We observed that increases in competency made it harder and harder for selection to pick the best genomes, because well-performing individuals could have done so either because their genomes were high-quality or because their high competency made up for a deficient genome. Specifically, the correlation between genotypic and phenotypic fitness drops to insignificant levels very rapidly (Figure 6C). This could be expected to result in complex dynamics because competency helps fitness of individuals but impairs the ability of the evolutionary process itself to capitalize on the hill-climbing search in fitness space by picking out the most elite genomes for mating in each generation (propagate improvements in genomes, which it can no longer clearly evaluate). Thus, we studied what happens when evolution also is allowed to control the degree of competency, which is biologically realistic as cellular capacities for sensing, computation, and action are themselves under evolutionary selection.

Surprisingly, we observed that the population does not, in the long term, drive towards maximizing competency (Figure 6A). It settles on a high value, but not one that is enough to produce a perfect phenotype from any genome whatsoever. When allowed to set the competency level freely, evolution produced phenotypically-perfect individuals but the genome remained slightly above 80% in quality, showing how competency and genome quality trade off at an equilibrium point (Figure 6B). We propose the following explanation.

Initially when evolution begins, a high gene value is required to create individuals with high fitness. At generation 40 or so, a high fitness is achieved by choosing high gene values over gene values. Post generation 50, the genotypic fitness has reached a stable value around 0.8. Theoretically, a competency value of 125 would be adequate to reach peak fitness from this point onwards. Because of this, evolution can get away by selecting a competency-gene greater than or equal to 125. Any value above 125 will ensure that peak fitness is reached. As a result, there is no selection pressure for evolution to pick the maximum possible competency value (400 in our system). There would be no advantage in choosing a competency value of 400 over 126. They are perceived as equal by the selection process. Further, there exists a random component to selection of gene values above 125. Two individuals, one with competency value 125 and another with competency value 400, will both have a phenotypic fitness of 1.0 (maximum). In our experiments, if >10% of the population have a fitness of 1.0 before selection, we pick the first 10%. This process of selection can lead to random competent genes being selected. Nonetheless they are well above 125, as seen in Fig.6A.

The paradox of robustness: why the animal with the worst genome has the best anatomy

The competency of a population can be seen as granting robustness against perturbations, i.e. competency resolves aberrations in the genome and lessens the burden on evolution. The role of robustness in evolution has been a popular topic of discussion [100-104]. A population's robustness is hypothesized to cause an evolutionary reduction in the adaptive performance of the population; a sort of maladaptation caused when improved robustness-traits layer on top of one another over evolutionary time and hide the underlying adaptive traits. This paradox has been shown to have broad implications on organismal design and is supposed to be a key aspect of evolution. Our results are in line with this paradox. Fig. 3 is a clear depiction of the role robustness plays in hindering the quality of the genome. At each generation, competency adds a layer of robustness which otherwise would have been exposed to evolution. Over evolutionary time, these robustness layers contribute collectively to worsen the genomes. When compared to a population with no competency (hardwired), competent genomes stabilize to a mediocre value whereas the untampered hardwired genomes rise steadily to maximum fitness. However, it must be noted that this is not necessarily a disadvantage. The paradox reveals the efficiency of Competency: Genomes need not be perfect; a stable value is all that is required for competency to boost an individual's fitness to maximum. Genetic information and problem-solving capacity of the cells work together to achieve a perfect solution to this fitness function.

There is a remarkable example of biology that is uniquely explained by the above dynamics[105]. Some species of planarian flatworms reproduce by fission and regeneration, which means that they exhibit somatic inheritance: any mutation that doesn't kill a cell is propagated into the next generation [87,106]. For hundreds of millions of years, these animals have accumulated genetic diversity and in fact single animals can be mixoploid [107,108], having different numbers of chromosomes. And yet, they regenerate perfectly after being cut into pieces, and do not age – they are champions of anatomical control. We propose that planaria are an example of runaway competency: when cells get really good at making up for genetic deficiencies, evolution has such a hard time selecting for the best genomes that further improvements concern the competency to reach their target morphology.

This is seen to some extent in other species; for example, human embryos are tolerant to being cut in half at early stages, creating monozygotic twins. However, in planaria the effect was apparently much stronger. We propose this also as an explanation for another curious aspect of planaria. In every other model species, mutant lines are available – fruit flies with different number of wings or color of eyes, mice with abnormal tails, and many more genetic strains that are available from stock centers. In planaria this does not exist – no morphologically-abnormal genetic strains have been reported. We suggest that this is because the cells have gotten so good at producing normal worms regardless of genetic changes, they automatically resist mutagenic effects that would normally found a new morphotype. In fact the only available abnormal line of planaria is a permanently two-headed form [109-111], which was produced not genetically but by manipulating bioelectrical signaling – the modality that is used to coordinate cellular competency [112-114], as is predicted by our model for species like planaria. Given their resistance to mutation, it's unclear how speciation in planaria happens, but it should be noted that the same bioelectrical strategy that controls computation and cognition (i.e., behavioral competencies) in brains has been shown to coax genetically wild-type planaria to grow the heads appropriate to other species [115,116].

Where the hard work is done: an intelligence ratchet

In planaria, most of the evolutionary "effort" seems to have gone in to perfecting the algorithm (the ability of cells to create a normal worm morphology), vs. keeping a clean genome, because of the vicious cycle of competency increases. We observed this in our model, asking where evolution was doing the work, once rising competency levels made it difficult to select the best genomes. We hypothesized that the process of evolution would spend more time optimizing the competency gene than any structural gene. To test this prediction, we tracked the amount of changes over evolutionary time that appear in "structural" genes vs. the competency gene (Figure 7), noting that the competency gene is changed significantly more often than any structural gene. When gains can no longer be efficiently made by tweaking the genome (once selection cannot reliably pick out the good genotypes), all the effort goes in to increasing the competency level.

This suggests the existence of a powerful ratchet mechanism in which evolution progressively becomes locked in to improvements in the intelligence of the agential material with which it works, with reduced pressure on the structural genes. A positive feedback loop in which evolution increasingly puts more effort into the developmental software than perfecting the hardware points to a possible drive for scaling intelligence in morphological and other spaces [37,39,117-120]. It is possible that a drive for increased competency is an ancient and ubiquitous pressure [121], which plays out to different degrees in different biological lineages based on other aspects of their environmental and reproductive hyper-parameters.

Conclusion

These results suggest a diverse research program in research on the evolutionary interplay between biological hardware (genomically-determined protein components in each cell) and software (multi-cellular physiological dynamics that control growth and form). We suggest that the field of basal cognition is an important part of understanding evolutionary developmental biology [57-59,77,78], and that intelligence (problem-solving competency) an evolutionary driver long before complex brains and muscle-driven behavior arise [36,37,40,122-129]. Beyond understanding natural evolution, we suggest that the design of autonomous robotics [130,131], synthetic life [132], and interventions for regenerative medicine [133] can all benefit from exploiting the multiscale competency architecture so richly exhibited by living forms.

**Acknowledgements**

We thank: Chris Fields, Steven Frank, Hananel Hazan, Santosh Manicka, and Richard Watson for helpful discussions, and Julia Poirier for assistance with the manuscript preparation. M.L. gratefully acknowledges support via John Templeton Foundation grant 62212.

# Figures and Legends

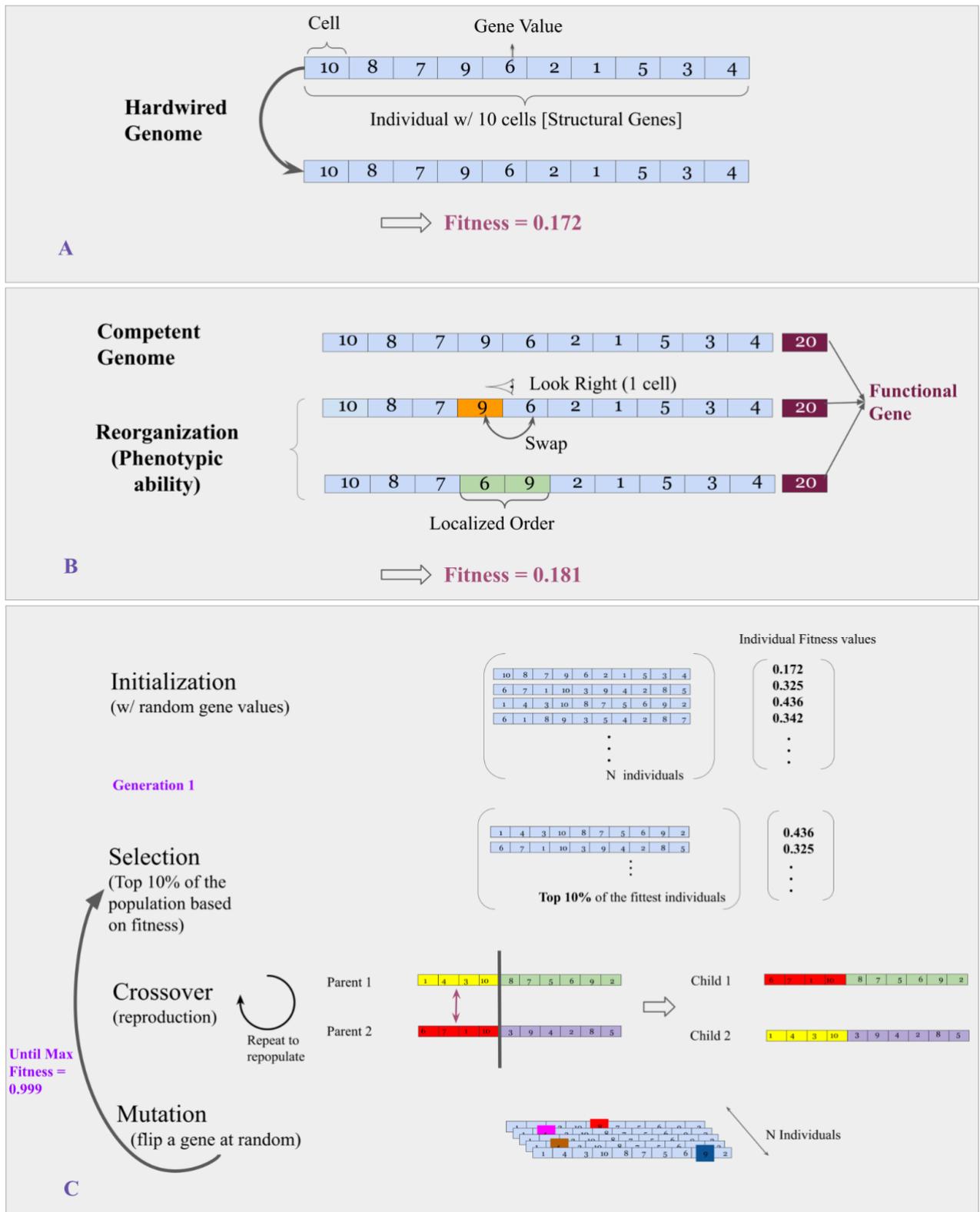

Figure 1: Schematic of experimental setup.

     A: Description of how a hardwired individual is defined: Each Hardwired individual is a 1-D array consisting of N cells (10 shown here). Each cell takes an integer value between [1, N], and is considered to be its "Structural Gene" (Genotype). The fitness of an individual depends on the degree

of order within its genes (0 implying no order and 1 implying ascending order). In the example shown, the individual is randomly initialized and hence has a fitness close to 0.

B: Description of how a competent individual is defined: Each competent individual is identical to a Hardwired individual except that it has the ability (phenotypic) to move its cells around (in a constrained manner) to achieve ordered arrangement (ascending). Each competent individual carries a functional gene. The functional gene can be locked down to a pre-specified value or can be made evolvable.

C: Description of the genetic algorithm used to evolve Hardwired and Competent individuals:

Step1- Populations of HW and Competent individuals are initialized with random cell values. Fitnesses of the HW individuals are computed. Competent individuals are allowed to rearrange their cells and fitnesses are calculated post-rearrangement.

Step2: Top 10% of the fittest individuals (genotypes) in both populations are selected.

Step3: The selected individuals are used to reproduce and re-populate their respective populations.

Step4: With a small probability, individuals of a population have a single cell-value (gene) flipped to a random number between [1, N]

Step 4 marks the end of a generation. A new generation begins once again with selection (step: 2). This cycle repeats until populations reach a fitness of 0.99 at which their genotypes would have near-perfect order.

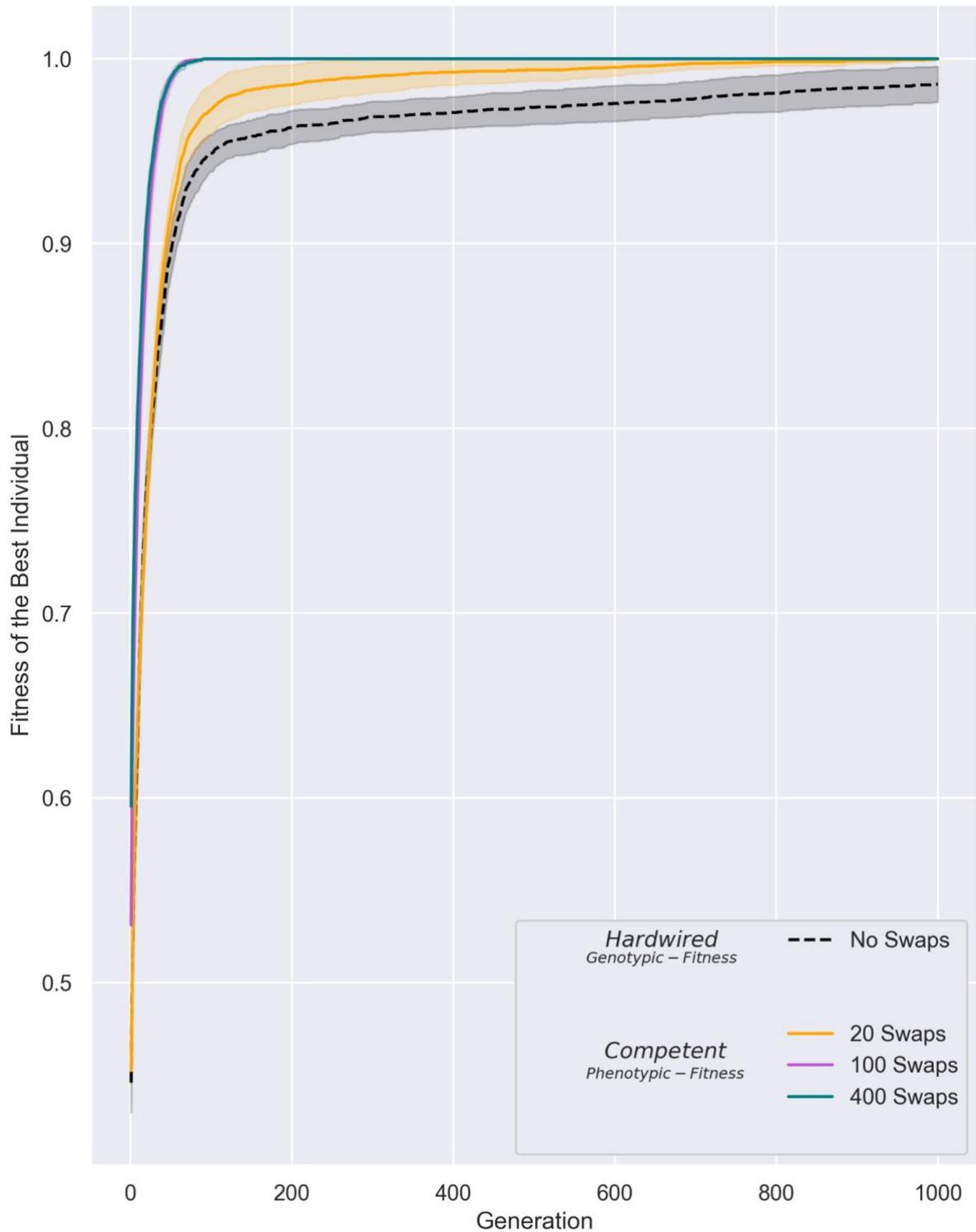

Figure 2. Competent individuals have a higher rate of fitness than their Hardwired counterparts.
In each generation the individual with the maximum fitness is shown. Competent populations are initialized with the ability to swap in a constrained manner (0/20/100/400 swaps per generation). Individuals with a higher competency level reach higher values of fitness faster. Shaded area depicts variance over 10 runs.

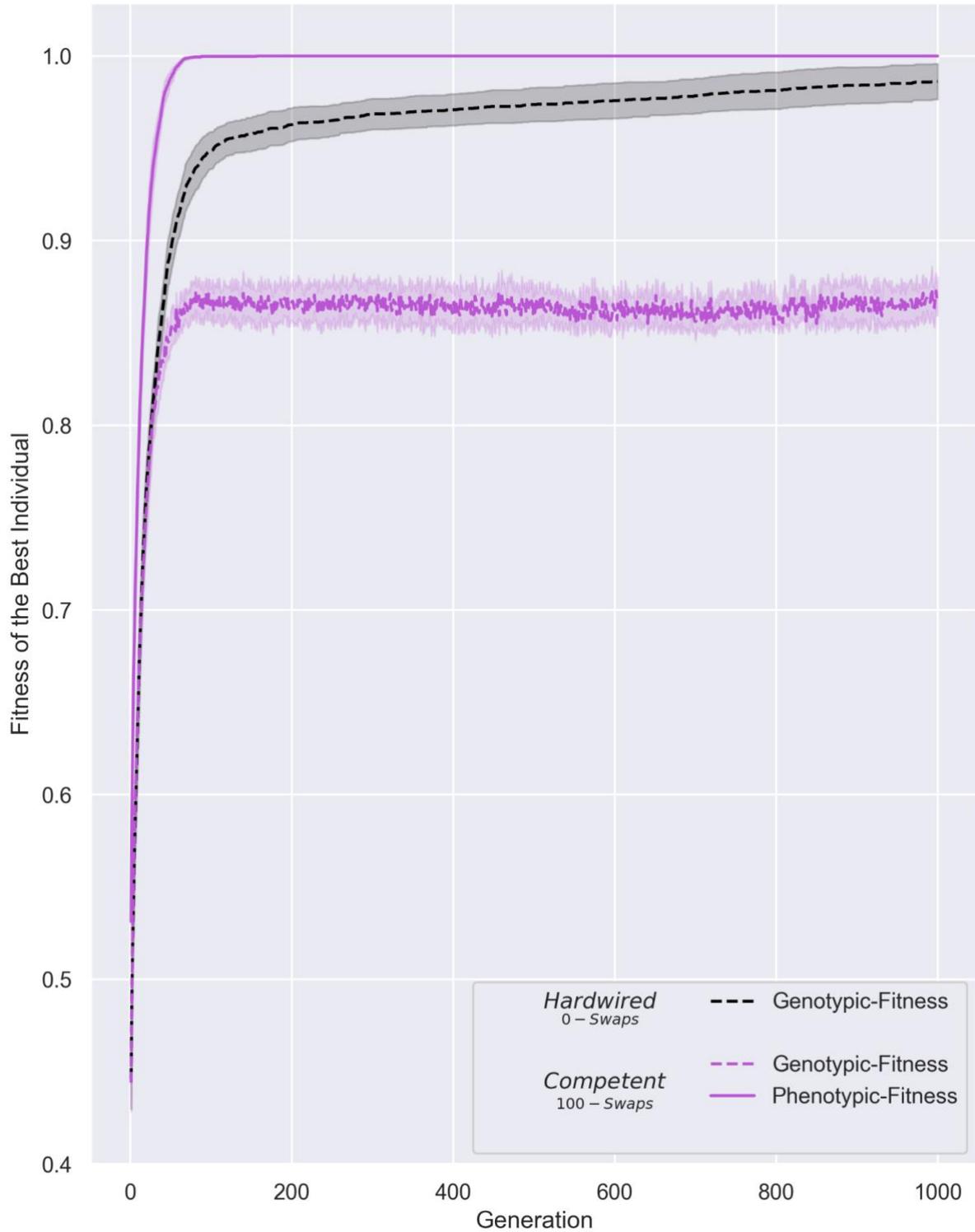

Figure 3. Competency comes at the expense of reduced genotypic fitness.
Fitnesses of the best individual in a Competent Population (with competency =100) is compared with that of a hardwired population over time. Shaded area represents variance over 10 runs.

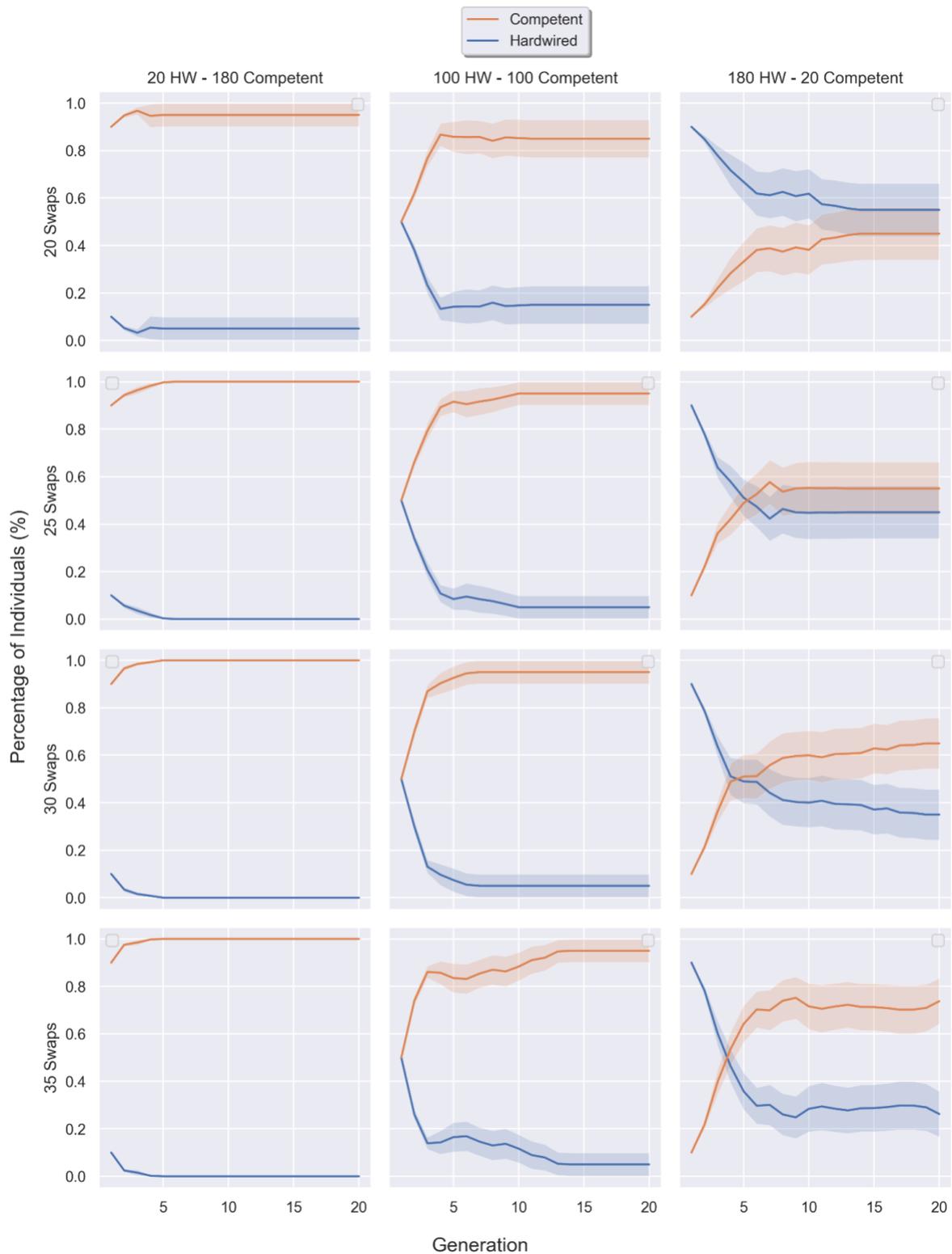

Figure 4: Competent individuals can dominate over hardwired individuals in a mixed setting only when given a minimum level of competency.

Prevalence of hardwired and competent individuals for different mixture proportions at various competency levels. Each column represents a different proportion of hardwired (HW) and Competent individuals mixed together for evolution within a hybrid population. Each row shows data from experiments at different competency levels (# of swaps permitted). Shaded area represents variance over 20 repeat runs of each experiment.

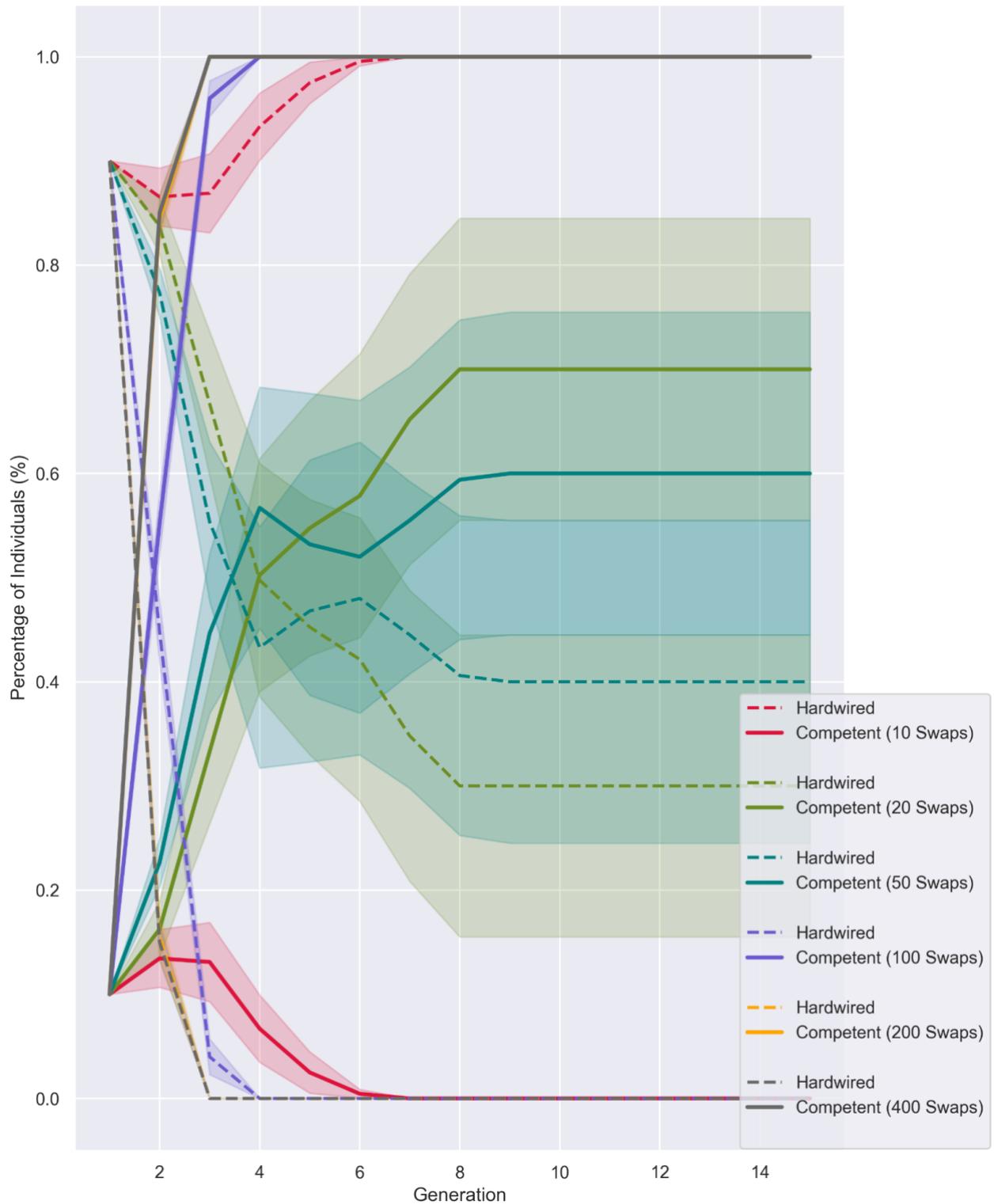

Figure 5: A competency of 100 or greater is required for the competent to drive hardwired to extinction when being a minority of the population.

Prevalence of a mixture of hardwired and competent individuals when competent individuals are a minority (180 hardwired to 20 competent) at different competency levels. Shaded area represents variance over 15 runs.

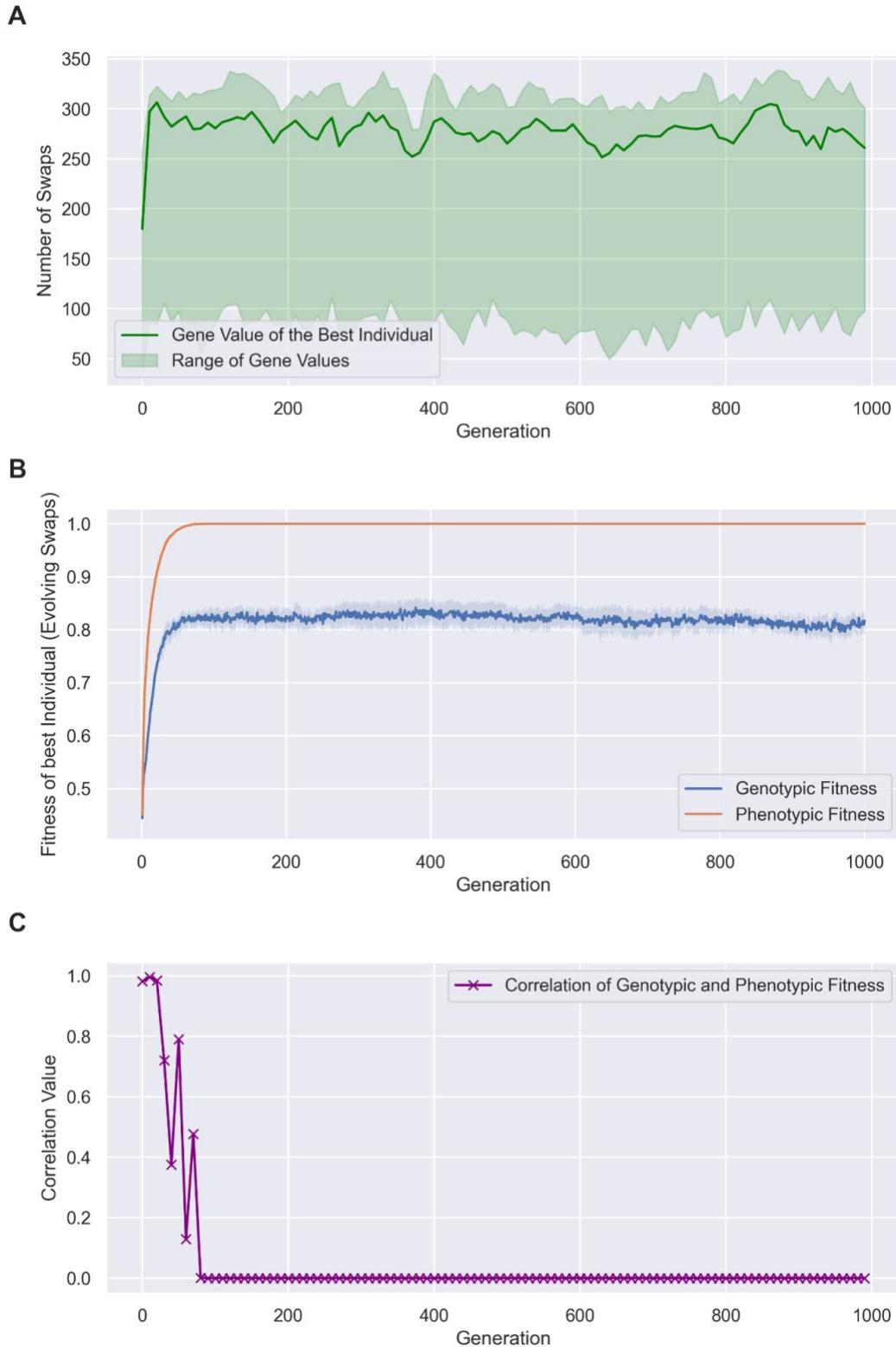

Figure 6: Competency does not require a perfect genome to boost fitness.
  A: Competency-gene value chosen over the course of evolution (shown as average values over sequences of 10 generations). Shaded area represents the range of competency values in the population.
  B: Fitness of the best individual in a population of competent individuals with evolvable competency. Shaded area represents variance over 10 runs. C: Correlation of the genotypic and phenotypic values of the population (shown as average values over sequences of 10 generations).

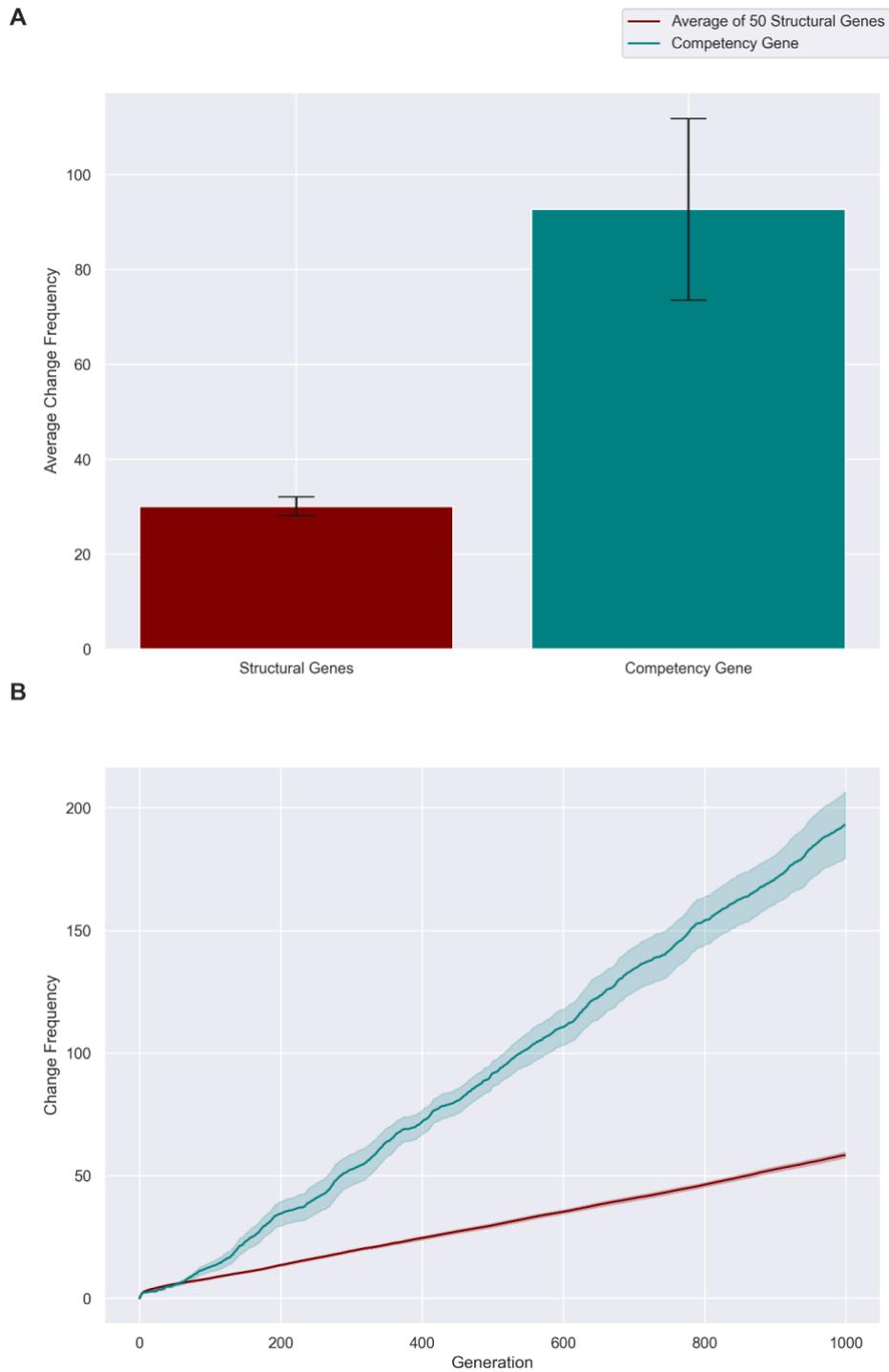

Figure 7: Evolution spends a greater proportion of time tweaking the competency gene compared to any structural gene

    A: Frequency of changes averaged over time. Error bars represent standard deviation over 10 runs. The competency gene is clearly modified at a greater rate per generation than any structural gene.

    B: Number of changes made to structural genes versus the competency gene over evolutionary time. Average value of change frequency in 50 structural genes are compared to those of the competency gene. The graph is cumulative in nature, i.e., the number of changes made in the previous generation carry forward to the next. Shaded area represents variance over 10 repeat runs.

**Tables**

| Competency Level \ Fitness | 0.65 | 0.75 | 0.80 | 0.90 | 0.97 | 1.00 |
|---|---|---|---|---|---|---|
| **No Competency (Hardwired)** | 9 | 18 | 25 | 52 | 359 | >1000 |
| **Level 20** | 9 | 19 | 25 | 44 | 92 | >1000 |
| **Level 100** | 3 | 8 | 11 | 21 | 38 | 156 |
| **Level 400** | 1 | 5 | 8 | 18 | 35 | 90 |

Table 1: The number of generations different competent populations take to break through a particular fitness-threshold. The table indicates the magnitude of assistance provided by competency to evolution.

| Competency Level \ Mixture Proportion | 20 HW: 180 Competent | 100 HW: 100 Competent | 180 HW : 20 Competent |
|---|---|---|---|
| **Level 20** | 0 | 0 | ∞ |
| **Level 25** | 0 | 0 | 5 |
| **Level 30** | 0 | 0 | 4 |
| **Level 35** | 0 | 0 | 3 |

Table 2: Details of how many generations competent individuals take to dominate (i.e., whenever a population is > 50% of the mixture) over hardwired individuals when mixed together in different proportions.